\documentclass[10pt,letterpaper]{article}
\usepackage{opex3}

\usepackage[latin9]{inputenc}
\usepackage{float}
\usepackage{bm}
\usepackage{amsmath}
\usepackage{amssymb}
\usepackage{esint}

\makeatletter
\usepackage{cite}

\renewcommand\maketitle{}
\DeclareMathAlphabet{\mathcal}{OMS}{cmsy}{m}{n}

\makeatother

\begin{document}

\title{Transformation inverse design}

\author{ David Liu,$^{1,*}$ Lucas H. Gabrielli,$^{2}$ Michal Lipson,$^{2,3}$,
and Steven G. Johnson$^{4}$}

\address{$^{1}$Department of Physics, MIT, Cambridge, MA 02139,
USA\\$^{2}$School of Electrical and Computer Engineering, Cornell
University, Ithaca, NY 14853, USA\\$^{3}$Kavli Institute, Cornell
University, Ithaca, NY 14853, USA\\$^{4}$Department of Mathematics,
MIT, Cambridge MA 02139, USA}\email{$^*$daveliu@mit.edu}
\begin{abstract}
We present a new technique for the design of transformation-optics
devices based on large-scale optimization to achieve the optimal effective
isotropic dielectric materials within prescribed index bounds, which
is computationally cheap because transformation optics circumvents
the need to solve Maxwell's equations at each step. We apply this
technique to the design of multimode waveguide bends (realized experimentally
in a previous paper) and mode squeezers, in which all modes are transported
equally without scattering. In addition to the optimization, a key
point is the identification of the correct boundary conditions to
ensure reflectionless coupling to untransformed regions while allowing
maximum flexibility in the optimization. Many previous authors in
transformation optics used a certain kind of quasiconformal map which
overconstrained the problem by requiring that the entire boundary
shape be specified \textit{a priori} while at the same time underconstraining
the problem by employing ``slipping'' boundary conditions that permit
unwanted interface reflections. 
\end{abstract}


\bibliographystyle{unsrt}

\section{Introduction}

\begin{figure}[t]
\centerline{\includegraphics[scale=0.17]{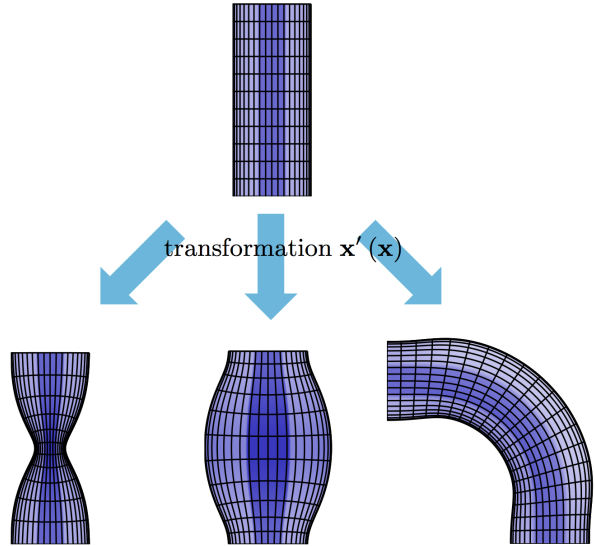}}

\caption{Three possible applications of transformation optics for multimode
waveguides: squeezer, expander, and bend. Dark areas indicate higher
refractive index. \label{fig:Three-possible-applications}}
\end{figure}

In this work, we introduce the technique of transformation inverse
design, which combines the elegance of transformation optics \cite{ref-wardpendry,ref-ulf_opticalconformalmapping,ref-pendryscience,ref-toreview,ref-nanoscale} (TO) with the power of large-scale optimization (inverse design),
enabling automatic discovery of the best possible transformation for
given design criteria and material constraints. We illustrate our
technique by designing multimode waveguide bends \cite{ref-landy,ref-curved,ref-strict_conformal,ref:conformal_exponential,ref-yao,ref-homogeneous,ref-roberts,ref-china_bend,ref-affine,ref-exp_bend,ref-arbitrary_bending,ref-adaptive,ref-confined-bend}
and mode squeezers \cite{ref-squeezing,ref-adaptive,ref-confined-bend,ref-china_squeeze,ref-ozgun},
then measuring their performance with finite element method (FEM)
simulations. Most designs in transformation optics use either hand-chosen
transformations \cite{ref-pendryscience,ref-china_bend,ref-affine,ref-macroscopic,ref-variable,ref-resistor,ref-controlscattering,ref-microwave,ref-macroscopic_science,ref-multifunctional,ref-inverse_to,ref-homogeneous,ref-squeezing,ref-roberts}
(which often require nearly unattainable anisotropic materials), or
quasiconformal and conformal maps \cite{ref-ulf_opticalconformalmapping,ref-lipendry,ref-light,ref-antenna,ref-lens,ref-3dquasi,ref-illusion,ref-extreme_angle,ref-groundcloak,ref-valentine_cloak,ref-gabrielli_nature_photon,ref-pendry-science-cloak,ref-freespace_cloak,ref-quasi-isotropic,ref-fiberchip,ref-platform,ref-planar_antenna,ref-squeezing,ref-landy,ref-strict_conformal,ref:conformal_exponential,ref-curved,ref-yao,ref-schwarz,ref-conformal_pennstate,ref-enhancing-image}
which can automatically generate nearly-isotropic transformations
(either by solving partial differential equations or by using grid
generation techniques) but still require\emph{ a priori }specification
of the entire boundary shape of the transformation. Further, neither
technique can directly incorporate refractive-index bounds. On the
other hand, most inverse design in photonics involves repeatedly solving
computationally expensive Maxwell equations for different designs
\cite{ref-frei-maxwell,ref-kao,ref-dobson-optimization,ref-vuckovic-design,ref-vuckovic-inverse,ref-topology,ref-zbend,ref-photon_nano_opt,ref-jensen-design,ref-topology_cloak,ref-topology_fab,ref-dobson-bandgap,ref-junction,ref-compensating,ref-levelset,ref-frei_waveguide,ref-tsuji,ref-derivatives-osher,ref-aperiodic,ref-top_opt_phbend}.
Transformation inverse design combines elements of both transformation
optics and inverse design while overcoming their limitations. First,
the use of optimization allows us to incorporate arbitrary fabrication
constraints while at the same time searching the correct space of
transformations without unnecessarily underconstraining or overconstraining
the problem. Second, instead of solving Maxwell's equations, we require
only simple derivatives to be computed at each optimization step.
This is because transformation optics works by using a coordinate
transformation~$\mathbf{x}^{\prime}(\mathbf{x})$ that warps light
in a desired way (e.g. mapping a straight waveguide to a bend, or
mapping an object to a point or the ground for cloaking applications
\cite{ref-lipendry,ref-pendryscience,ref-groundcloak,ref-gabrielli_nature_photon,ref-nanorods,ref-valentine_cloak,ref-pendry-science-cloak,ref-euclid,ref-leonhardt_book})
and then employing transformed materials which are given in terms
of the Jacobian~$\bm{\mathcal{J}}_{ij}=\partial x_{j}^{\prime}/\partial x_{i}$
to mathematically mimic the effect of the coordinate transformation.
This transforms all solutions of Maxwell's equations in the same way
(as opposed to non-TO multimode devices which often have limited bandwidth
and/or do not preserve relative phase between modes \cite{ref-fan_bends,ref-hightrans,ref-jensen-design,ref-rightangle,ref-vic_liu,ref-broadband_phbend,ref-noda_highlyconfined,ref-compact_flexible,ref-top_opt_phbend,ref-yetanother_bend,ref-achromatic}),
and is therefore particularly attractive for designing multimode optical
devices \cite{ref-schmiele,ref-elements,ref-finite_embedded,ref-light,ref-adaptive}
(such as mode squeezers, expanders, splitters, couplers, and multimode
bends) with no intermodal scattering. (Before TO was first used for cloaking, electrostatic cloaking using anisotropic conductivities \cite{ref-calderon,ref-eit} had been proposed). Examples of such transformations
are shown in Fig.~\ref{fig:Three-possible-applications}. For a more comprehensive treatment of the physical and mathematical ideas of transformation media used in this paper, see \cite{ref-leonhardt_book}.

One major difficulty with transformation optics is that most functions~$\mathbf{x}^{\prime}(\mathbf{x})$
yield highly anisotropic and magnetic materials. In principle, these
transformed designs can be fabricated with anisotropic microstructures
\cite{ref-gradient_index,ref-pendry_meta,ref-birefringent,ref-microwave}
or naturally birefringent materials \cite{ref-macroscopic,ref-macroscopic_science}.
However, in the infrared regime (where metals are lossy) it is far
easier to instead fabricate effectively isotropic dielectric materials,
provided that the refractive index falls within the given bounds~$n_{\mathrm{min}}$
and $n_{\mathrm{max}}$ of the fabrication process (for example, subwavelength
nanostructures \cite{ref-gabrielli_nature_photon,ref-valentine_cloak,ref-nanorods,ref-pendry-science-cloak,ref-lens,ref-gradient_index,ref-graded-index-photonic,ref-graded_photon_nano,ref-prl-polarize,ref-platform}
or waveguides with variable thickness \cite{ref-grayscale,ref-graded,ref-ulrich,ref-zernike,ref-marom,ref-brazas}).
This requirement means that we would prefer to consider the subset
of transformations that can be mapped to approximately isotropic dielectric
materials. 

The theory of transformation optics with nearly isotropic materials
is intimately connected to the subjects of conformal maps (which are
isotropic by definition \cite{ref-ulf_opticalconformalmapping,ref-rudin,ref-shilov}) and \emph{quasiconformal
maps} {[}which in mathematical analysis are defined as \emph{any}
orientation-preserving transformation with bounded anisotropy (as
quantified in Sec.~\ref{sub:Nearly-isotropic-transformations}){]}.
However, in transformation optics the term ``quasiconformal'' has
become confusingly associated with only a single choice of quasiconformal
map suggested by Li and Pendry \cite{ref-lipendry}. In that work,
Li and Pendry proposed minimizing a mean anisotropy with ``slipping''
boundary conditions (defined in Sec.~\ref{sub:Nearly-isotropic-transformations}),
which turns out to yield a transformation that is essentially conformal
up to a \emph{constant stretching} (and thus anisotropy) everywhere.
This map, which also happens to minimize the peak anisotropy given
the slipping boundary conditions \cite{ref-lipendry,ref-lipendry-private},
is sometimes confusingly called ``the quasiconformal map'' \cite{ref-landy,ref-3dquasi,ref-extreme_angle,ref-valentine_cloak,ref-pendry-science-cloak}.
However, we point out in Sec.~\ref{sub:Nearly-isotropic-transformations}
that slipping boundary conditions are not the correct choice if one
wishes to ensure a reflectionless interface between transformed and
untransformed regions. Instead, for interfaces to be reflectionless
requires at least continuity of the transformation~$\mathbf{x}^{\prime}$
at the interface \cite{ref-yan_arxiv,ref-luzi,ref-yan-cloak,ref-finite_embedded}
and, as we show in Sec.~\ref{sub:Conformal-maps} for the case of
isotropic dielectric media, continuity of the Jacobian~$\mathcal{J}$
as well. If one fixes the transformation on part or all of the boundary
(instead of just the corners) and minimizes the peak anisotropy, the
result is called (in analysis) an \emph{extremal} quasiconformal map
\cite{ref-extremal,ref-elliptic,ref-handbook_teich,ref-geometric-handbook,ref-acta,ref-quasi}.
We point out in Sec.~\ref{sub:Isotropic-transformations} that this
extremal quasiconformal map can never be conformal except in trivial
cases. Additionally, previous work in quasiconformal transformation
optics underconstrained the space of transformations in one way but
overconstrained it in another. Li and Pendry's method, along with
other work on extremal quasiconformal maps in mathematical analysis,
assumed that the entire boundary \emph{shape} of the transformed domain
is specified \emph{a priori} (even if the \emph{value} of the transformation
at the boundary is not specified). In contrast, transformation inverse
design allows parts of the boundary shape to be freely chosen by the
optimization, only fixing aspects of the boundary that are determined
by the underlying problem (e.g. the input/output facets of the boundary
in Fig.~\ref{fig:Three-possible-applications}) as explained in Secs.~\ref{sub:Generalized-bend-transformation},
allowing a much larger space of transformations to be searched. Also,
for such stricter boundary conditions, minimizing the mean anisotropy
is \emph{not} equivalent to minimizing the peak anisotropy \cite{ref-elliptic,ref-france,ref-deformations,ref-heisenberg},
and we argue below that the peak anisotropy is a better figure of
merit for transformation optics in general.

We solve all of these problems by using large-scale numerical optimization
to find the transformation with minimal peak anisotropy that exactly
obeys continuity conditions at the boundary with untransformed regions.
This allows the input/output interfaces to transition smoothly and
continuously into untransformed devices while also satisfying fabrication
constraints (e.g. bounds on the attainable refractive indices and
bend radiii). A large space of arbitrary smoothly varying transformations
(that satisfy the continuity conditions and fabrication constraints)
is explored quickly and efficiently by parametrizing in a ``spectral''
basis \cite{boyd_cheb,boyd} of Fourier harmonics and Chebyshev polynomials.
The optimized transformation is then scalarized (as in the case of
previous work on quasiconformal transformation optics) into an isotropic
dielectric material that guides modes with minimal intermodal scattering
and loss. In the case of a multimode bend, for which our design was
recently fabricated and characterized \cite{ref-dliu}, we achieve
intermodal scattering at least an order of magnitude smaller than
a conventional non-TO bend.

In Sec.~\ref{sec:Transformation-optics}, we review the equations
of transformation optics. In Sec.~\ref{sub:Isotropic-transformations},
we describe situations where the transformation-designed material
can be mapped to isotropic media. In Sec.~\ref{sub:Conformal-maps},
we point out that such isotropic transformations, due to their analyticity,
always have undesirable interface discontinuities when coupled into
untransformed regions. In Secs.~\ref{sub:Nearly-isotropic-transformations}
and \ref{sub:Scalarization-errors-for}, we review the techniques
of quasiconformal mapping (as used in both the transformation optics
and mathematical analysis literature) and scalarization of nearly
isotropic transformations. We show that the inherent restrictions
of quasiconformal mapping can be circumvented by directly optimizing
the map using transformation inverse design. In Secs.~\ref{sub:Simple-circular-bend}
and \ref{sub:Generalized-bend-transformation}, we design a nearly
isotropic transformation for a $90^{\circ}$-bend by perturbing from
the highly anisotropic circular bend transformation. In Secs.~\ref{sub:Setup-of-optimization}
and \ref{sub:Spectral-parameterization}, we set up the bend optimization
problem and the spectral parameterization. In Sec.~\ref{sub: results},
we present the optimized structure, which reduces anisotropy by several
orders of magnitude compared to the circular TO bend. In Sec.~\ref{sub:Peak-minimized-structure},
we present finite element simulation results comparing our optimized
design to the conventional non-TO bend and the circular TO bend. In
Secs.~\ref{sub:minmax} we show that minimizing the mean anisotropy
can lead to pockets of high anisotropy (which in turn leads to greater
intermodal scattering) while minimizing the peak does not. In Secs.~\ref{sub:Tradeoff-between-anisotropy},
we discuss the tradeoff between the bend radius and the optimized
anisotropy. In Sec.~\ref{sec:Mode-squeezer} we briefly present methods
and results for applying transformation inverse design to optimize
mode squeezers.

\section{Mathematical preliminaries}

\subsection{Transformation optics \label{sec:Transformation-optics}}

The frequency domain Maxwell equations (fields $\sim e^{-i\omega t}$),
without sources or currents, in linear isotropic dielectric media
{[}$\bm{\varepsilon}=\varepsilon(\mathbf{x}),\,\bm{\mu}=\mu_{0}${]}
are
\begin{align}
\nabla\times\mathbf{H} & =-i\omega\varepsilon(\mathbf{x})\mathbf{E}\nonumber \\
\nabla\times\mathbf{E} & =i\omega\mu_{0}\mathbf{H}.\label{eq:maxwell}
\end{align}
Consider a coordinate transformation~$\mathbf{x}^{\prime}\left(\mathbf{x}\right)$
with Jacobian~$\mathcal{J}_{ij}=\frac{\partial x_{j}^{\prime}}{\partial x_{i}}$.
We define the \emph{primed} gradient vector as $\nabla^{\prime}\equiv\left(\frac{\partial}{\partial x^{\prime}},\,\frac{\partial}{\partial y^{\prime}},\,\frac{\partial}{\partial z^{\prime}}\right)=\bm{\mathcal{J}}^{-1}\nabla$
and the primed fields as $\mathbf{E}^{\prime}\equiv\bm{\mathcal{J}}^{-1}\mathbf{E}$
and $\mathbf{H}^{\prime}\equiv\bm{\mathcal{J}}^{-1}\mathbf{H}$. One
can then rewrite Eq.~(\ref{eq:maxwell}), after some rearrangement
\cite{ref-wardpendry,ref-leonhardt_review}, as
\begin{align}
\nabla^{\prime}\times\mathbf{H}^{\prime} & =-i\omega\bm{\varepsilon}^{\prime}\mathbf{E}^{\prime}\nonumber \\
\nabla^{\prime}\times\mathbf{E}^{\prime} & =i\omega\bm{\mu}^{\prime}\mathbf{H^{\prime}},\label{eq:primed_maxwell}
\end{align}
where the effects of the coordinate transformation have been mapped
to the equivalent tensor materials
\begin{equation}
\bm{\mu}^{\prime}=\mu_{0}\frac{\mathcal{J}^{T}\hspace{-2bp}\mathcal{J}}{\det\bm{\mathcal{J}}}\qquad\qquad\bm{\varepsilon}^{\prime}=\varepsilon(\mathbf{x})\frac{\mathcal{J}^{T}\hspace{-2bp}\mathcal{J}}{\det\bm{\mathcal{J}}}.\label{eq:transformed_materials}
\end{equation}
This equivalence has become known as \emph{transformation optics}
(TO). It is actually the specific case of a much more general result from general relativity \cite{ref-jp_generalrel}. For a further discussion of space--time transformations and connections to negative refraction, see \cite{ref-leonhardt_book} and \cite{ref-leonhardt_genrel}.

\begin{figure}
\centerline{\includegraphics[scale=0.15]{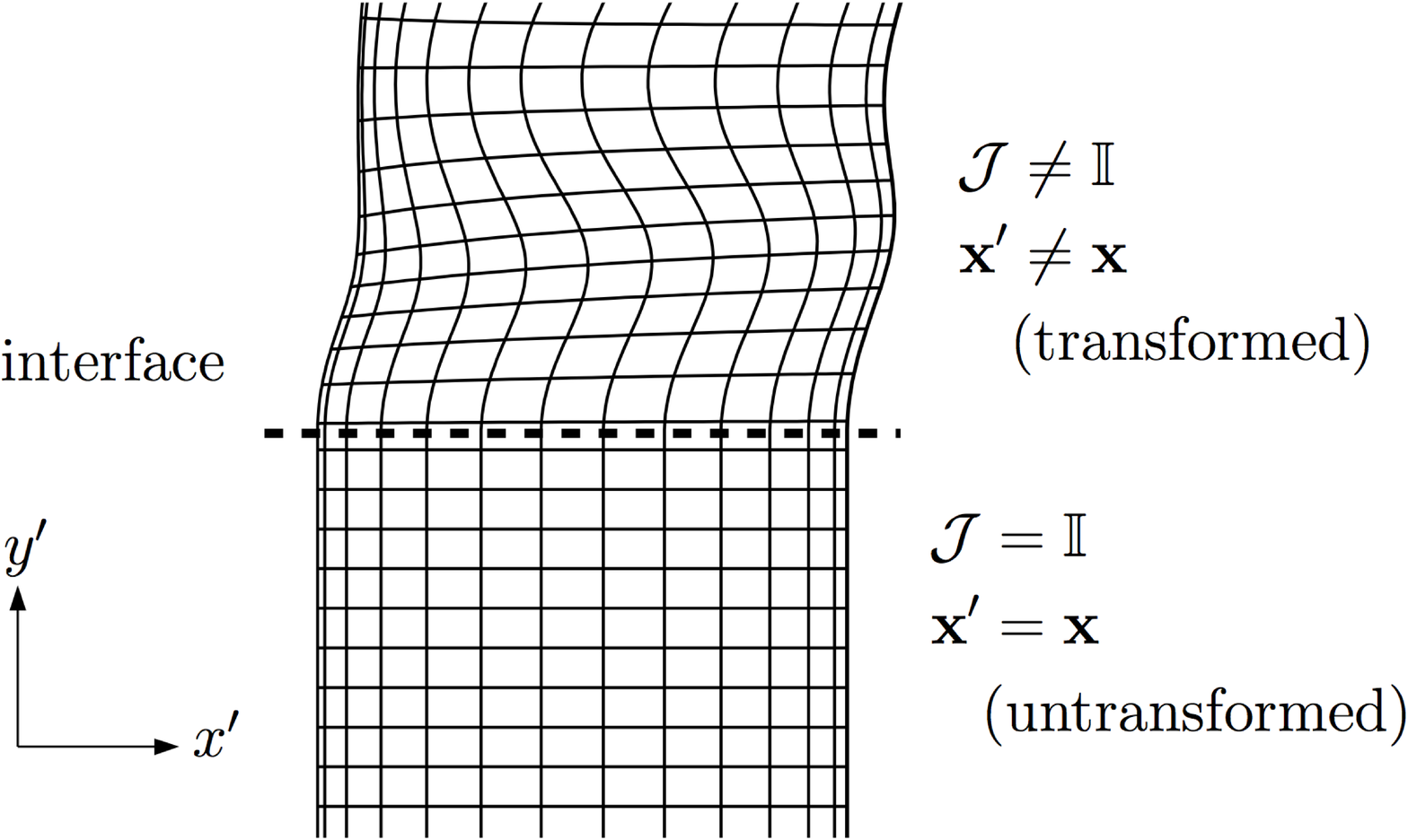}}

\caption{The interface between the transformed and untransformed region must
have $\mathbf{x}^{\prime}$ continuous in order for there not to be
any interface reflections.\label{fig:Jboundary}}
\end{figure}

Most useful applications of TO require that the transformation be
coupled to untransformed regions (e.g. the input and output straight
waveguides in the case of a bend transformation, or the surrounding
air region for the case of a ground-plane cloaking transformation).
However, in order for TO to guarantee that the interface between transformed
and untransformed regions be reflectionless, the transformation must
be equivalent to a \emph{continuous }transformation of all space that
is the identity~$\mathbf{x}^{\prime}(\mathbf{x})=\mathbf{x}$ in
the ``untransformed'' regions, as depicted in Fig.~\ref{fig:Jboundary}.
More generally, the untransformed regions can be simple rotations
or translations, but when examining a particular interface, we can
always choose the coordinates to be $\mathbf{x}^{\prime}=\mathbf{x}$
at that interface. It is clear by construction that continuous $\mathbf{x}^{\prime}$
is sufficient for reflectionless interfaces \cite{ref-luzi,ref-yan-cloak,ref-finite_embedded},
and this is in fact a necessary condition as well \cite{ref-yan_arxiv}.
Although a general anisotropic transformation need only have $\mathbf{x}^{\prime}(\mathbf{x})$
continuous at the interface , we show below that an \emph{isotropic}
transformation will also have a continuous $\bm{\mathcal{J}}$ at
the interface. These boundary conditions are essential for designing
useful transformations without interface reflections. 

\subsection{Transformations to isotropic dielectric materials \label{sub:Isotropic-transformations}}

For the vast majority of transformations, the materials in Eq.~(\ref{eq:transformed_materials})
are anisotropic tensors. However, for certain transformations, the
tensors are effectively scalar. Suppose that the transformation~$\mathbf{x}^{\prime}(\mathbf{x})$
is 2D ($z^{\prime}=z$ and $\frac{\partial\mathbf{x}^{\prime}}{\partial z}=0$),
making $\bm{\mathcal{J}}$ block-diagonal (with the $zz$ element
independent of the $xy$ block). Then, the $xy$ block of $\mathcal{J}^{T}\hspace{-2bp}\mathcal{J}$
is isotropic if and only if the diagonal elements are equal and the
off-diagonal elements vanish: 
\begin{align}
\left|\nabla x^{\prime}\right|^{2}-\left|\nabla y^{\prime}\right|^{2} & =0\nonumber \\
\nabla x^{\prime}\cdot\nabla y^{\prime} & =0.\label{eq:isotropic equations}
\end{align}
In this case, the $\bm{\mathcal{J}}$ part of Eq.~(\ref{eq:transformed_materials})
becomes 
\begin{equation}
\frac{\mathcal{J}^{T}\hspace{-2bp}\mathcal{J}}{\det\bm{\mathcal{J}}}=\begin{pmatrix}1\\
 & 1\\
 &  & \frac{1}{\det\bm{\mathcal{J}}}
\end{pmatrix}.\label{eq:isotropicJJT}
\end{equation}

This isotropy has different implications for transverse-magnetic (TM)
polarized modes in 2D (which have $\mathbf{E}=E\hat{z}$ and $\mathbf{H}\cdot\hat{z}=0$)
versus transverse-electric (TE) polarized modes (which have $\mathbf{E}\cdot\hat{z}=0$
and $\mathbf{H}=H\hat{z}$). For TM-polarized modes, the fields~$\mathbf{E}^{\prime},\,\mathbf{H}^{\prime}$
in the primed coordinate system are also TM-polarized and Eq.~(\ref{eq:primed_maxwell})
becomes
\begin{align*}
\nabla^{\prime}\times\mathbf{H}^{\prime} & =-i\omega\frac{\varepsilon(\mathbf{x})}{\det\bm{\mathcal{J}}}\mathbf{E}^{\prime}\\
\nabla^{\prime}\times\mathbf{E}^{\prime} & =i\omega\mu_{0}\mathbf{H^{\prime}}.
\end{align*}
Hence, for TM modes, an isotropic transformation can be exactly mapped
to an isotropic dielectric material. Similarly, for TE-polarized modes
the equivalent material is isotropic and \emph{magnetic}. However,
if $\det\bm{\mathcal{J}}$ varies slowly compared to the wavelengths
of the fields, then the transformation can \emph{approximately} still
be mapped to an isotropic dielectric material by making an eikonal
approximation (as in \cite{ref-jackson} Ch.~8.10) and commuting
$1/\mu^{\prime}$ with one of the curls in the Maxwell equations.
In particular, Eq.~(\ref{eq:primed_maxwell}) can be written:
\[
\nabla^{\prime}\times\nabla^{\prime}\times\mathbf{E}^{\prime}=\omega^{2}\varepsilon_{0}\mu_{0}(\det\bm{\mathcal{J}})\mathbf{E}^{\prime}+\mathcal{O}(\nabla\det\bm{\mathcal{J}}),
\]
where the last term can be neglected for slowly varying transformations.
Because the TM case is conceptually simpler and does not require this
extra approximation, we work with it exclusively for the rest of this
paper. Also, because the non-trivial aspects of the transformation
occur in the $xy$ plane, we hereafter use $\mathcal{J}$ to denote
the $xy$ block.

\subsection{Conformal maps and uniqueness\label{sub:Conformal-maps}}

If a transformation has an isotropic \textbf{$\mathcal{J}$}, then
the transformation preserves angles in the $xy$ plane. Additionally,
if $\det\mathcal{J}>0$, then the transformation also preserves handness
and orientation. The combination of these two properties is called
a conformal map \cite{ref-rudin}, and is the only case where the
situation in Sec.~\ref{sub:Isotropic-transformations} can be realized.
We only consider transformations with $\det\mathcal{J}>0$ in order
to restrict ourselves to dielectric materials. Also, a $\det\mathcal{J}>0$
transformation coupled continuously to an untransformed ($\det\mathcal{J}=1$)
region would require singularities ($\det\mathcal{J}=0$) at some
points. Conformal maps are described by analytic functions, which
are of the form $x^{\prime}+iy^{\prime}=w^{\prime}(w)$ (where $w\equiv x+iy$
is the untransformed complex coordinate) and whose real and imaginary
parts satisfy the Cauchy--Riemann equations of complex analysis \cite{ref-rudin,ref-shilov}.

However, true conformal maps cannot directly be used for transformation
optics in typical applications, because of the impossibility of coupling
them to untransformed regions with the boundary conditions discussed
in Sec.~\ref{sec:Transformation-optics}. In particular, the uniqueness
theorem of analytic functions \cite[Thm.~10.39]{ref-shilov} tells
us that if $w^{\prime}(w)=w$ in some region, then $w^{\prime}(w)=w$
everywhere (similarly for a simple rotation or translation in some
regions). 

As a corollary, in the limit where a transformation becomes more and
more isotropic in the neighborhood of an interface, it must have a
continuous $\mathcal{J}$, not just a continuous $\mathbf{x}'(\mathbf{x})$.
It is easy to see this explicitly in the example of Fig.~\ref{fig:Jboundary}:
continuity of $\mathbf{x}'(\mathbf{x})$ at the interface requires
that $\frac{\partial\mathbf{x}^{\prime}}{\partial x}=(1,\,0)$ on
both sides of the interface, which determines the first row of $\mathcal{J}$.
The isotropy of $\mathcal{J}^{T}\hspace{-2bp}\mathcal{J}$ then forces
$\mathcal{J}=\mathbb{I}$. Therefore, in the sections that follow
(where we search for \emph{approximately} isotropic maps), we will
impose the condition of continuous $\mathcal{J}$ as a boundary condition
on our transformations. The resulting transformations are nearly isotropic
in the interior and exactly isotropic on the interfaces. This condition,
discussed at the end of Sec.~\ref{sub:Scalarization-errors-for},
also has the useful consequence of producing a continuous refractive
index~$n^{\prime}=\sqrt{\varepsilon^{\prime}\mu^{\prime}}$.

\subsection{Quasiconformal maps and measures of anisotropy \label{sub:Nearly-isotropic-transformations}}

Because true conformal maps cannot be used, one widely used alternative
is to search for a nearly\emph{ }isotropic transformation, which can
be \emph{approximated }by an isotropic material at the cost of some
scattering corrections to the exactly transformed modes of the nearly
isotropic material. To do this, one must first quantify the measure
of anisotropy that is to be minimized. The isotropy condition of Eq.~(\ref{eq:isotropic equations})
is equivalent to $\lambda_{1}=\lambda_{2}$, where $\lambda_{1}(x,\, y)\geq\lambda_{2}(x,\, y)$
are the two eigenvalues of $\mathcal{J}^{T}\hspace{-2bp}\mathcal{J}$.
While $\lambda_{1}-\lambda_{2}$ works as a measure of anisotropy,
it is convenient for optimization purposes to define \emph{differentiable
}measures that can be expressed directly in terms of the trace and
determinant of $\mathcal{J}$, and precisely such quantities have
been developed in the literature on quasiconformal maps \cite{ref-quasi,ref-ahlfors,ref-elliptic,ref-surface_quasi,ref-extremal}. 

A general transformation is an arbitrary function of $x$ and $y$
or, equivalently, an arbitrary function $w^{\prime}(w,\,\bar{w})$,
of $w$ and $\bar{w}=x-iy$, which may not be analytic in $w$. The
anisotropy can be related to the \emph{Beltrami coefficient }\cite{ref-quasi,ref-elliptic}
\[
\mu_{\mathrm{B}}\left(w,\,\bar{w}\right)\equiv\left(\frac{\partial w^{\prime}}{\partial\bar{w}}\right)\left(\frac{\partial w^{\prime}}{\partial w}\right)^{-1}.
\]
The term \emph{quasiconformal map }refers to \emph{any }map that has
bounded $\left|\mu_{\mathrm{B}}\right|<1$, which includes all non-singular
sense preserving ($\det\mathcal{J}>0$) transformations. It can be
shown that the \emph{linear distortion}~$K$ \cite{ref-extremal,ref-elliptic}
satisfies
\[
K\equiv\frac{1+\left|\mu_{\mathrm{B}}\right|^{2}}{1-\left|\mu_{\mathrm{B}}\right|^{2}}=\sqrt{\frac{\lambda_{1}}{\lambda_{2}}}\geq1.
\]
Various other measures of anisotropy have been defined in the grid
generation literature \cite{cite-handbook_grid,ref-fundamentalsgrid},
including the \emph{Winslow} and \emph{Modified Liao} functionals,
which are given by $\Phi\equiv\int\mathrm{d}^{2}x\,(K+\frac{1}{K})$
and $\Phi\equiv\int\mathrm{d}^{2}x\,(K^{2}+\frac{1}{K^{2}})$, respectively.
However, for the rest of this work we refer to the quantity~$\mathbb{K}-1\geq0$
as the ``anisotropy'', where $\mathbb{K}$ is the \emph{distortion
function }\cite{ref-extremal,ref-elliptic}, defined as
\begin{equation}
\mathbb{K}\left(x,\, y\right)\equiv\frac{1}{2}\left(K+\frac{1}{K}\right)=\frac{\mathrm{tr}\mathcal{J}^{T}\hspace{-2bp}\mathcal{J}}{2\det\mathcal{J}}\geq1.\label{eq:kminusone}
\end{equation}
The tensor~$\frac{\mathcal{J}^{T}\hspace{-2bp}\mathcal{J}}{\det\mathcal{J}}$
is known as the \emph{distortion tensor} \cite{ref-elliptic}. 

As mentioned in the introduction, an \emph{extremal} quasiconformal
map is one that minimizes the peak anisotropy, given the shape of
the transformed region and the values of the transformation on some
or all of the boundary \cite{ref-extremal,ref-ahlfors,ref-quasi,ref-elliptic}.
Because $\mathbb{K}$, $\mathbb{K}-1$, $K$, and $\left|\mu_{\mathrm{B}}\right|$
are all monotonic functions of one another, they are equivalent for
the purpose of finding an extremal quasiconformal map. However, $\mathbb{K}$
is numerically convenient because it is a differentiable function
of the entries of $\mathcal{J}$. These quantities are \emph{not}
generally equivalent for minimizing the \emph{mean} anisotropy \cite{ref-elliptic,ref-france,ref-deformations,ref-heisenberg},
and we argue in Sec.~\ref{sub:minmax} that the peak anisotropy is
a better figure of merit. However, in the special case where the value
of the transformation is only fixed at the \emph{corners} of the domain
and is allowed to vary freely in between (a ``slipping''' boundary
condition), Li and Pendry showed that it is equivalent to minimize
the mean (either Winslow or modified-Liao) or the peak anisotropy,
and that these yield a constant-anisotropy map (a uniform scaling
of a conformal map) \cite{ref-lipendry-private} that they and other
authors have used for transformation optics \cite{ref-light,ref-antenna,ref-lens,ref-3dquasi,ref-illusion,ref-extreme_angle,ref-groundcloak,ref-valentine_cloak,ref-gabrielli_nature_photon,ref-pendry-science-cloak,ref-freespace_cloak,ref-quasi-isotropic,ref-fiberchip,ref-platform,ref-planar_antenna,ref-squeezing,ref-landy}.
However, the slipping boundary will generally lead to reflections
at the interface between the transformed and untransformed regions
because of the resulting discontinuity in the transformation, which
can only be reduced by making the transformation domain very large
in cases (e.g. cloaking) with localized deformations. In order to
design compact transformation-optics devices, especially for applications
such as bends where the deformation is nonlocalized, we will instead
impose continuity of the transformation and/or its Jacobian on the
input/output facets of the domain, while at the same time allowing
the shape of some or all of the boundary to vary (unlike all previous
work on quasiconformal maps, to our knowledge).

\subsection{Scalarization errors for nearly isotropic materials \label{sub:Scalarization-errors-for}}

The minimum-anisotropy quasiconformal map is then \emph{scalarized}
(as in \cite{ref-lipendry}) by approximating it with an isotropic
dielectric material. As shown in Sec.~\ref{sub:Isotropic-transformations},
a perfectly isotropic 2D transformation of a geometry with an isotropic
dielectric material that guides TM modes~$\mathbf{E}_{0},\,\mathbf{H}_{0}$
can be mapped to a transformed material and geometry that is also
isotropic dielectric and guides TM modes~$\mathbf{E}_{0}^{\prime},\,\mathbf{H}_{0}^{\prime}$.
This is exact for $\mathbb{K}=1$, but for a \emph{nearly} isotropic
transformation with $\mathbb{K}>1$, the equivalent permeability is
$\bm{\mu}^{\prime}=\mathbb{I}+\Delta\bm{\mu}$, where the anisotropic
part~$\Delta\bm{\mu}$ is proportional to $\mathbb{K}-1$ to lowest
order. While $\Delta\bm{\mu}^{\prime}\neq0$ cannot be fabricated
using dielectric gradient index processes, one can neglect this small
correction so that the actual fabricated material has permeability~$\bm{\mu}_{\mathrm{approx}}^{\prime}=\mathbb{I}$.
In practice, we absorb any $\Delta\bm{\mu}^{\prime}$ into $\varepsilon^{\prime}$
by multiplying $\varepsilon^{\prime}$ by the average eigenvalue of
$\bm{\mu}^{\prime}$
\begin{equation}
\left\langle \bm{\mu}^{\prime}\right\rangle =\frac{\lambda_{1}+\lambda_{2}}{2\sqrt{\lambda_{1}\lambda_{2}}}=\frac{\mathrm{tr}\mathcal{J}^{T}\hspace{-2bp}\mathcal{J}}{2\det\mathcal{J}}\label{eq:average_eigenvalue}
\end{equation}
 but this does not change the $\mathcal{O}\left(\mathbb{K}-1\right)$
error. 

A Born approximation \cite{ref-sgj_roughness,ref-snyder,ref-chew}
tells us that, given an exact transformation with no scattering, any
small error of $\Delta\bm{\varepsilon}$ and $\Delta\bm{\mu}$ will
generically lead to scattered fields with magnitudes of $\mathcal{O}(\left|\Delta\bm{\varepsilon}\right|+\left|\Delta\bm{\mu}\right|)$
and scattered power of $\mathcal{O}(\left|\Delta\bm{\varepsilon}\right|^{2}+\left|\Delta\bm{\mu}\right|^{2})$.
The modes of the approximate scalarized material~$\bm{\mu}_{\mathrm{approx}}^{\prime},\,\varepsilon_{\mathrm{approx}}^{\prime}$
are then the exact guided modes plus scattered power corrections of
$\mathcal{O}(|\Delta\bm{\mu}|^{2})=\mathcal{O}[(\mathbb{K}-1)^{2}]$.

A similar analysis explains why we must explicitly impose continuity
of $\mathcal{J}$ at the input/output facets of the domain. As explained
in Sec.~\ref{sub:Conformal-maps}, a purely isotropic transformation
in the neighborhood of the interface, along with a continuity of $\mathbf{x}^{\prime}$,
would automatically yield continuous $\mathcal{J}$, so one might
hope that minimizing anisotropy would suffice to obtain a nearly continuous
$\mathcal{J}$. Unfortunately, as we show in the Appendix, the resulting
discontinuity in $\det\mathcal{J}$ (and hence the discontinuity in
the refractive index) is of order $\mathcal{O}(\sqrt{\mathbb{K}-1})$,
which would lead to $\mathcal{O}(\mathbb{K}-1)$ power loss due to
reflections, much larger than the $\mathcal{O}[\left(\mathbb{K}-1\right)^{2}]$
power scattering from anisotropy in the interior. This would make
it pointless to minimize the anisotropy in the interior, since the
boundary reflections would dominate. In fact, our initial implementation
of the bend optimization in Sec.~\ref{sub:Spectral-parameterization}
did not enforce continuity of $\mathcal{J}$, and we obtained a large
$2\%$ index discontinuity at the endfacets for $\max_{\mathbf{x}}\mathbb{K}-1\approx0.0005$.
Therefore, in Sec.~\ref{sub:Spectral-parameterization} we impose
continuity of $\mathcal{J}$ explicitly.

\subsection{General optimization of anisotropy\label{sub:Numerical-optimization-of}}

In this paper, we directly minimize $\mathbb{K}$ using large-scale
numerical optimization while keeping track of constraints on the transformation
$\mathbf{x}^{\prime}$ and its Jacobian $\mathcal{J}$, as well as
the engineering fabrication bounds $n_{\mathrm{min}}$ and $n_{\mathrm{max}}$.
By using numerical optimization, we can in principle achieve both
a lower mean anisotropy and a lower peak anisotropy than by traditional
quasiconformal mapping, since the optimization is also free to vary
the boundary shape (with at most the input/output interfaces fixed,
although in some cases their locations and shapes are allowed to vary
as well). The minimization problem can be written, for example, as
\begin{equation}
\min_{\mathbf{x}^{\prime}(\mathbf{x})}\left\Vert \mathbb{K}(\mathbf{x})\right\Vert \;\mathrm{subject\, to}\;\begin{cases}
\mathbf{x}^{\prime},\,\mathcal{J}\,\mathrm{continuous\, at\, input/output\, interfaces}\\
n_{\mathrm{min}}\le n^{\prime}(\mathbf{x})\le n_{\mathrm{max}}
\end{cases},\label{eq:general_min}
\end{equation}
where $\left\Vert \mathbb{K}(\mathbf{x})\right\Vert $ is a \emph{functional}
norm taken over the domain of $\mathbf{x}^{\prime}(\mathbf{x})$.
We consider two possible norms: the $L_{1}$ norm (the mean $\left\langle \mathbb{K}\right\rangle _{\mathbf{x}}$),
and the $L_{\infty}$ norm ($\max_{\mathbf{x}}\mathbb{K}$). We show
in Sec.~\ref{sub:minmax} that minimizing the mean can lead to pockets
of high anisotropy which can cause increased scattering. Directly
optimizing the peak anisotropy on the other hand, avoids such pockets
while simultaneously keeping the mean nearly as low. The continuity
of $\mathbf{x}^{\prime}$ and $\mathcal{J}$ at the input/output interfaces,
as well as other constraints on the interface locations, are imposed
implicitly by the parametrization of $\mathbf{x}^{\prime}(\mathbf{x})$
(as explained in Sec.~\ref{sub:Spectral-parameterization}).

\section{Multimode Bend design \label{sec:perturbations}}

In this section, we design a bend transformation (depicted in Fig.~\ref{fig:schematic})
using general methods to (locally) solve the optimization problem
of Eq.~(\ref{eq:general_min}). In contrast, previous work on TO bend
design either utilized materials that were either anisotropic or consisted
of multiple stacked isotropic layers \cite{ref-homogeneous,ref-roberts,ref-china_bend,ref-affine,ref-exp_bend,ref-adaptive,ref-confined-bend}
or employed slipping boundary conditions \cite{ref-landy,ref-curved,ref-strict_conformal,ref:conformal_exponential,ref-yao}
(which result in endfacet reflections when coupled to untransformed
waveguide).

\subsection{Simple circular bends \label{sub:Simple-circular-bend}}

First, we consider a simple circular bend transformation (which we
refer to hereafter as the circular TO bend) that maps a rectangular
segment of length $L$ and width unity (in arbitrary distance units
to be determined later) into a bend with inner radius $R$ and outer
radius $R+1$ (as shown in Fig.~\ref{fig:schematic}). For convenience,
we choose the untransformed coordinates to be $R\le x\le R+1$ and
$-\frac{L}{2}\le y\le\frac{L}{2}$, with the untransformed segment
length~$L=\frac{\pi R}{2}$ equal to the inner arclength of the bend.
The transformation~$\mathbf{x}^{\prime}(\mathbf{x})$ can be written
as
\begin{align}
x^{\prime} & =r\,\cos\theta\nonumber \\
y^{\prime} & =r\,\sin\theta\nonumber \\
z^{\prime} & =z,\label{eq:polar}
\end{align}
where $r=x$ and $\theta=\frac{y}{R}$. While $\mathbf{x}^{\prime}$
is continuous at the input/output interfaces $y=\pm\frac{L}{2}$,
one issue is that $\mathcal{J}$ is not continuous there, which can
be seen from $\det\mathcal{J}=\frac{x}{R}\neq1$. Another issue is
that $\bm{\mu}^{\prime}\neq\mu_{0}\mathbb{I}$ is highly anisotropic.
The anisotropy for this transformation is $\mathcal{\mathbb{K}}(x,\, y)-1=\frac{x}{2R}+\frac{R}{2x}-1$,
which has a peak value $\max_{\mathbf{x}}\mathbb{K}-1\approx\frac{1}{2R^{2}}$
for $R\gg1$ at the outer radius $x=R+1$. Note that one can instead
choose $r=\exp(\frac{\pi x}{2L})$, which gives the \emph{conformal}
bend~$x^{\prime}+iy^{\prime}=\exp[\frac{\pi}{2L}(x+iy)]$. As explained
in Sec.~\ref{sub:Conformal-maps}, this map has zero anisotropy,
but neither $\mathbf{x}^{\prime}$ nor $\bm{\mathcal{J}}$ are continuous
at the input/output interfaces, leading to large reflections there.

\subsection{Generalized bend transformations \label{sub:Generalized-bend-transformation}}

In order to address the problems of the circular TO bend, we look
for minimum anisotropy \emph{and }continuous-interface transformations
of the form of Eq.~(\ref{eq:polar}), where the intermediate polar
coordinates are now arbitrary functions~$r(x,\, y)$ and $\theta(x,\, y)$.
The ratio~$L/R$ is now an optimization parameter. The Jacobian then
satisfies
\begin{align*}
\mathrm{\mathrm{tr}}\mathcal{J}^{T}\hspace{-2bp}\mathcal{J} & =\left|\nabla r\right|^{2}+\left|r\nabla\theta\right|^{2}\\
\det\mathcal{J} & =\left|\nabla r\times r\nabla\theta\right|.
\end{align*}
We find that the optimization always seems to prefer a symmetric bend
(and if the optimum is unique, it must be symmetric), so we impose
a mirror symmetry in order to halve our search space: 
\begin{align}
r(x,\, y) & =r(x,\,-y)\nonumber \\
\theta(x,\, y) & =-\theta(x,\,-y).\label{eq:symmetry}
\end{align}
We also require interface continuity of $\mathbf{x}^{\prime}$ and
$\mathcal{J}$ (as discussed in Sec.~\ref{sub:Conformal-maps}), which
give the conditions at $y=\pm\frac{L}{2}$:
\begin{align}
r & =x\nonumber \\
\theta & =\pm\frac{\pi}{4}\nonumber \\
\frac{\partial r}{\partial y} & =0\nonumber \\
\frac{\partial\theta}{\partial y} & =\frac{1}{x}.\label{eq:continuity}
\end{align}

\subsection{Numerical optimization problem\label{sub:Setup-of-optimization}}

Besides minimizing the objective function~$\mathbb{K}$, the optimization
must keep track of several constraints. First, any fabrication method
will bound the overall refractive index~$n^{\prime}$ to lie between
some values~$n_{\mathrm{min}}$ and $n_{\mathrm{max}}$. We choose
units so that the width of the transformed region is unity ($R\leq x\leq R+1$),
and consider transforming a straight waveguide of width~$\Delta_{\mathrm{w}}<1$.
$\Delta_{\mathrm{w}}$ should be small enough so that the exponential
tails of the waveguide modes are negligible outside the transformed
region. In the straight waveguide segment to be transformed (as well
as the straight waveguides to be coupled into the input and output
interfaces of the bend), $n(\mathbf{x})$ is high in the core~$\left|x-R-\frac{1}{2}\right|<\frac{\Delta_{\mathrm{w}}}{2}$
and low in the cladding~$\left|x-R-\frac{1}{2}\right|>\frac{\Delta_{\mathrm{w}}}{2}$.
For convenience, we write this refractive index as a product~$n(\mathbf{x})=n_{0}p(x)$
of an overall refractive index~$n_{0}$ and a normalized profile~$p(x)$
that is unity in the cladding and some value greater than unity in
the core (determined by the ratio of the high and low index regions
of the straight waveguide). The transformed refractive index is given
by
\[
n^{\prime}\left(\mathbf{x}\right)=\sqrt{\varepsilon^{\prime}\mu^{\prime}}=n_{0}p\left(x\right)\sqrt{\frac{\mathrm{tr}\mathcal{J}^{T}\hspace{-2bp}\mathcal{J}}{2\left(\det\mathcal{J}\right)^{2}}}
\]
where the average eigenvalue~$\mu^{\prime}$ of the magnetic permeability
Eq.~(\ref{eq:average_eigenvalue}) has been absorbed into the dielectric
index. The overall refractive-index scaling~$n_{0}$ is then allowed
to freely vary as a parameter in the optimization. Second, like the
circular TO bend, the optimum TO bend is expected to have a tradeoff
between the bend radius and anisotropy. Because of this expected tradeoff,
we can choose to either minimize $R$ while keeping $\mathbb{K}$
fixed, or minimize $\mathbb{K}$ while keeping $R$ fixed. We focus
on the latter choice, since the bend radius is the more intuitive
target quantity to know beforehand. Also, we find empirically that
optimizing $\mathbb{K}$ converges much faster than optimizing $R$
while yielding the same local minima.

With these constraints, there are several ways to set up the optimization
problem, depending on which norm we are minimizing. One method is
to minimize the peak anisotropy~$\max_{\mathbf{x}}\mathbb{K}$ with
$\mathbf{x}\in G$ for some grid~$G$ of some points to be defined
in Sec.~\ref{sub:Spectral-parameterization}. However, the peak (the
$L_{\infty}$ norm) is not a differentiable function of the design
parameters, so it should not be directly used as the objective function.
Instead, we perform a standard transformation \cite{boyd}: we introduce
a dummy variable~$t$ and indirectly minimize the peak~$\mathbb{K}$
using a differentiable inequality constraint between $t$ and $\mathbb{K}(\mathbf{x})$
at all $\mathbf{x}\in G$: 
\begin{equation}
\min_{r(\mathbf{x}),\,\theta(\mathbf{x}),\, n_{0},\, L,\, t}t\;\mathrm{subject\, to}:\;\begin{cases}
\mathrm{continuity\; conditions}\;\,\ref{eq:symmetry},\,\ref{eq:continuity}\\
n_{\mathrm{min}}\le n_{0}p\left(x\right)\sqrt{\frac{\mathrm{tr}\mathcal{J}^{T}\hspace{-2bp}\mathcal{J}}{2\left(\det\mathcal{J}\right)^{2}}}\le n_{\mathrm{max}}\,\mathrm{for\,}\mathbf{x}\in G\\
R=R_{0}\\
\mathcal{\mathbb{K}}(\mathbf{x})\le t\,\mathrm{for\,}\mathbf{x}\in G
\end{cases}.\label{eq:min_t method}
\end{equation}
For comparison, we explain in Sec.~\ref{sub:minmax} why the $L_{\infty}$
norm is better to minimize than the $L_{1}$ norm (the mean anisotropy).

The minimization of the $L_{1}$ norm, $\left\langle \mathbb{K}\right\rangle _{\mathbf{x}}=\int\mathbb{K}\,\mathrm{d}x\,\mathrm{d}y/\mathrm{area}$
{[}which \emph{is} differentiable in terms of the parameters~$r(\mathbf{x})$,
$\theta(\mathbf{x})$, $n_{0}$, and $L${]} is implemented as 
\begin{equation}
\min_{r(\mathbf{x}),\,\theta(\mathbf{x}),\, n_{0},\, L}\left\langle \mathbb{K}\right\rangle _{\mathbf{x}}\;\mathrm{subject\, to}:\;\begin{cases}
\mathrm{continuity\; conditions}\;\,\ref{eq:symmetry},\,\ref{eq:continuity}\\
n_{\mathrm{min}}\le n_{0}p\left(x\right)\sqrt{\frac{\mathrm{tr}\mathcal{J}^{T}\hspace{-2bp}\mathcal{J}}{2\left(\det\mathcal{J}\right)^{2}}}\le n_{\mathrm{max}}\;\mathrm{for\,}\mathbf{x}\in G\\
R=R_{0}
\end{cases}.\label{eq:min_mean}
\end{equation}

\begin{figure}[t]
\centerline{\includegraphics[scale=0.13]{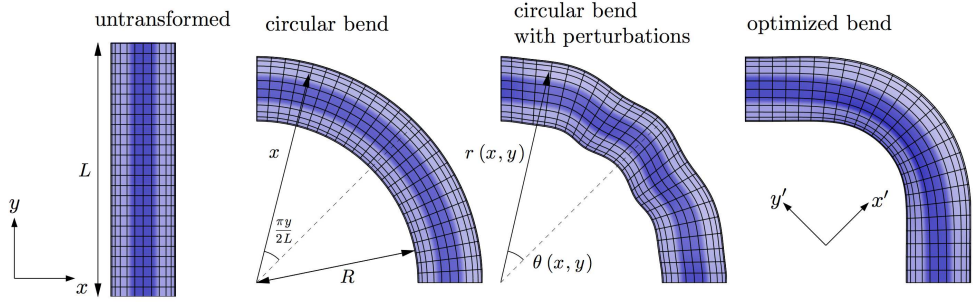}}

\caption{In the transformation process, the untransformed straight waveguide
is bent, perturbed, and optimized. Darker regions indicate higher
refractive index \label{fig:schematic}}
\end{figure}

We use the circular bend~$r=x,\,\theta=\frac{\pi y}{2L}=\frac{y}{R}$
as a starting guess, and search the space of general transformations~$r(\mathbf{x}),\,\theta(\mathbf{x})$
by perturbing from this base case. (We only perform \emph{local }optimization;
not global optimization, but comment in Sec.~\ref{sub:Tradeoff-between-anisotropy}
on a simple technique to avoid being trapped in poor local minima.)
As explained in Sec.~\ref{sub:Spectral-parameterization}, the perturbations
will be parameterized such that the symmetry and continuity constraints
are satisfied automatically. Figure~\ref{fig:schematic} shows a schematic
of the bend transformation optimization process. First, the straight
region is mapped to a circular bend. Then, the intermediate polar
coordinates~$r$ and $\theta$ for every point~$\mathbf{x}$ are
perturbed, using an optimization algorithm described at the end of
Sec.~\ref{sub:Spectral-parameterization}, and the desired norm (either
$L_{1}$ and $L_{\infty}$) of the anisotropy is computed. This process
is repeated at each optimization step until the structure converges
to a local minimum in $\left\Vert \mathcal{\mathbb{K}}\right\Vert $.

\subsection{Spectral parameterization\label{sub:Spectral-parameterization}}

To faciliate efficient computation of the objective and constraints,
the functions~$r$ and $\theta$ can be written as the circular bend
transformation plus perturbations parametrized in the spectral basis
\cite{boyd_cheb,boyd}:

\begin{align}
r\left(x,\, y\right) & =x+\sum_{\ell,\, m}^{N_{\ell},\, N_{m}}C_{\ell m}^{r}T_{\ell}(2x-2R-1)\,\cos\frac{2m\pi y}{L}\nonumber \\
\theta\left(x,\, y\right) & =\frac{\pi y}{2L}+\frac{1}{x}\sum_{\ell,\, m}^{N_{\ell},\, N_{m}}C_{\ell m}^{\theta}T_{\ell}(2x-2R-1)\,\sin\frac{2m\pi y}{L},\label{eq:perturbations}
\end{align}
where the coordinate~$2x-2R-1$ has been centered appropriately for
the domain~$\left[-1,\,1\right]$ of degree-$\ell$ Chebyshev polynomials
$T_{\ell}$. The sines and cosines have been chosen to satisfy the
mirror-symmetry conditions of Eq.~(\ref{eq:symmetry}). The sine series
also automatically satisfies the second continuity condition of Eq.~(\ref{eq:continuity}).
In order to satisfy the rest of the conditions, the following constraints
are also imposed:
\begin{align}
\sum_{m}^{N_{m}}C_{\ell m}^{r}\left(-1\right)^{m} & =0\nonumber \\
\sum_{m}^{N_{m}}C_{\ell m}^{\theta}\left(-1\right)^{m}m & =\begin{cases}
\frac{L}{8\pi}-\frac{R}{4}-\frac{1}{8}, & \ell=0\\
-\frac{1}{8}, & \ell=1\\
0, & \ell\geq2
\end{cases}.\label{eq:cconstraint}
\end{align}
These equations are solved to simply eliminate the $C_{\ell N_{m}}^{r,\,\theta}$
coefficients before optimization. 

This spectral parametrization has several advantages over finite-element
discretizations such as the piecewise-linear parameterization of \cite{ref-extremal}.
First, the spectral basis converges exponentially for smooth functions
\cite{boyd_cheb}. We found that only a small number ($N_{\ell}\times N_{m}<100$)
of spectral coefficients~$C^{r,\,\theta}$ are needed to achieve
very low-anisotropy ($\mathcal{\mathbb{K}}-1\approx10^{-4}$) transformations.
Second, if the fabrication process favors slowly varying transformations
(or if these are needed to make the eikonal approximation for the
TE polarization, as in Sec.~\ref{sub:Isotropic-transformations}),
this constraint may be imposed simply by using smaller $N_{\ell}$
and $N_{m}$. 

With this spectral parameterization, the formulation of the optimization
problem Eq.~(\ref{eq:min_t method}) becomes 
\begin{equation}
\min_{\left\{ C_{\ell m}^{r,\theta}\right\} ,\, n_{0},\, L,\, t}t\;\mathrm{subject\, to}:\;\begin{cases}
\mathrm{constraint\,\ref{eq:cconstraint}}\\
n_{\mathrm{min}}\le n_{0}p\left(x\right)\sqrt{\frac{\mathrm{tr}\mathcal{J}^{T}\hspace{-2bp}\mathcal{J}}{2\left(\det\mathcal{J}\right)^{2}}}\le n_{\mathrm{max}}\,\mathrm{for\,}\mathbf{x}\in G\\
R=R_{0}\\
\mathbb{K}(\mathbf{x})\le t\,\mathrm{for\,}\mathbf{x}\in G
\end{cases}.\label{eq:min_t coefficients}
\end{equation}

The local optimization was performed using the derivative-free COBYLA
non-linear optimization algorithm \cite{powell_cobyla,cite-powell_review}
in the NLopt package \cite{nlopt}. In principle, we can make the
optimization faster by analytically computing the derivatives of the
objective and constraints with respect to the design parameters and
using a gradient-based optimization algorithm, but that is not necessary
because both $\mathrm{tr}\mathcal{J}^{T}\hspace{-2bp}\mathcal{J}$
and $\det\mathcal{J}$, which determine all the non-trivial objective
and constraint functions in this optimization problem, are so computationally
inexpensive to evaluate that the convergence rate is not a practical
concern.

\section{Optimization results \label{sub: results}}

\subsection{Minimal peak anisotropy\label{sub:Peak-minimized-structure}}

\begin{figure}[t]
\centerline{\includegraphics[scale=0.15]{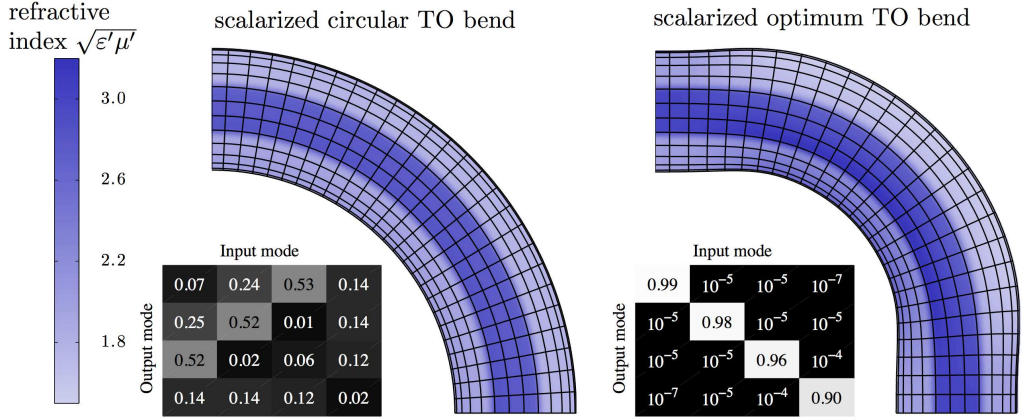}}\caption{Optimization decreases anisotropy by a factor of $10^{-4}$, while
dramatically improving the scattered-power matrix. \label{fig:Mreduced}}
\end{figure}

A $\min\left\Vert \mathbb{K}\right\Vert _{\infty}$ design is shown
in Fig.~\ref{fig:Mreduced}, along with the scalarized circular TO
bend for comparison. The bend radius was $R=2$ and the number of
spectral coefficients was $N_{\ell}=5,\, N_{m}=8$. The objective
and constraints were evaluated on a $100\times140$ grid $G$ in $\mathbf{x}$
(Chebyshev points in the $x$ direction and a uniform grid in the
$y$ direction). This design had $\max_{\mathbf{x}}\mathbb{K}-1\approx5\times10^{-4}$
and mean $\left\langle \mathbb{K}\right\rangle -1\approx10^{-4}$.
In comparison, the circular TO bend of the same radius has $\max_{\mathbf{x}}\mathcal{\mathbb{K}}-1\approx0.1$
and $\left\langle \mathbb{K}\right\rangle -1\approx10^{-2}$. 

The $R=2$ optimized design structure was compared in finite-element
Maxwell simulations (using the FEniCS code \cite{ref-fenics}), with
the conventional non-TO bend {[}simply bending the waveguide profile
around a circular arc with $n^{\prime}(\mathbf{x}^{\prime})=n(\mathbf{x})${]}
and the scalarized circular TO bend. The four lowest-frequency modes
of a multimode straight waveguide were injected at the input interface~$y=\frac{L}{2}$,
and the scattered-power matrix~$T$ was computed using the measured
fields at the output interface $y=-\frac{L}{2}$. The scattered-power
matrix is defined as
\[
T_{ij}=\left|\intop_{R}^{R+1}\mathrm{d}x\,\hat{\theta}\cdot\left(\mathbf{E}_{j}^{0}\times\mathbf{H}_{i}\right)\right|_{y=-\frac{L}{2}}^{2},
\]
where $-\hat{\theta}$ is the propagation direction of the guided
modes, $\mathbf{E}_{j}^{0}$ is the normalized electric field of the
$j$th exactly guided mode of the non-scalarized material $\left(\bm{\mu}^{\prime},\,\varepsilon^{\prime}\right)$,
and $\mathbf{H}_{i}$ is the actual magnetic field of the approximate
scalarized material at the interface after injecting a normalized
mode~$\mathbf{E}_{i}^{0}$ at the input interface. This makes $T_{ij}$
equal to the power scattered into the $j$th output mode from the
$i$th input mode. For a straight waveguide, which has no intermodal
scattering, $T=\mathbb{I}$. Figure~\ref{fig:Mreduced} shows a dramatically
improved $T$ for the scalarized and optimized TO bend compared to
the scalarized circular TO bend. {[}The rows and columns of $T$ for
the circular bend add up to less than one because some power has either
been scattered out of the waveguide entirely, or some power has been
scattered into fifth or higher-order modes. The rows and columns of
$T$ for the optimized bend add up to nearly 1, with the small deficiency
due to the $\mathcal{O}\left(\mathbb{K}-1\right)$ out-of-bend and
higher-order intermodal scattering as well as mesh-descretization
error.{]}

\begin{figure}[t]
\centerline{\includegraphics[scale=0.12]{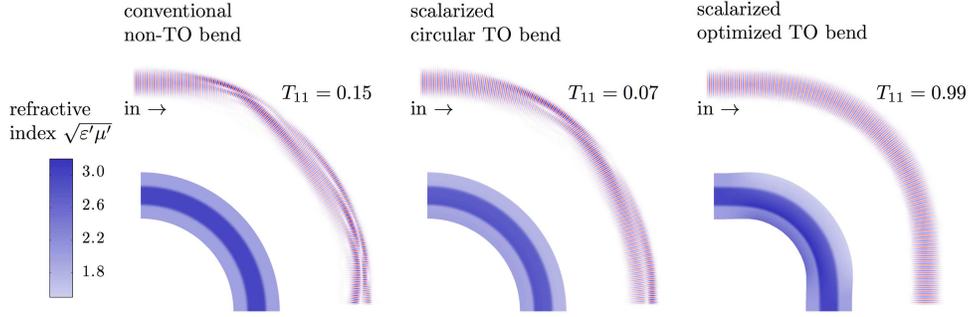}}

\caption{FEM field profiles show heavy scattering in the conventional non-TO
and scalarized circular bends, but very little scattering in the optimized
bend. \label{fig:Finite-element-simulations}}
\end{figure}

The electric-field profiles for the fundamental mode, displayed in
Fig.~\ref{fig:Finite-element-simulations}, show a dramatic difference
in the performance of the optimized structure versus the other structures.
Both the conventional and circular TO bend show heavy intermodal scattering
in the bend region, while the optimized transformation displays very
little scattering. 

\subsection{Minimizing max versus minimizing mean\label{sub:minmax}}

\begin{figure}[t]
\centerline{\includegraphics[scale=0.15]{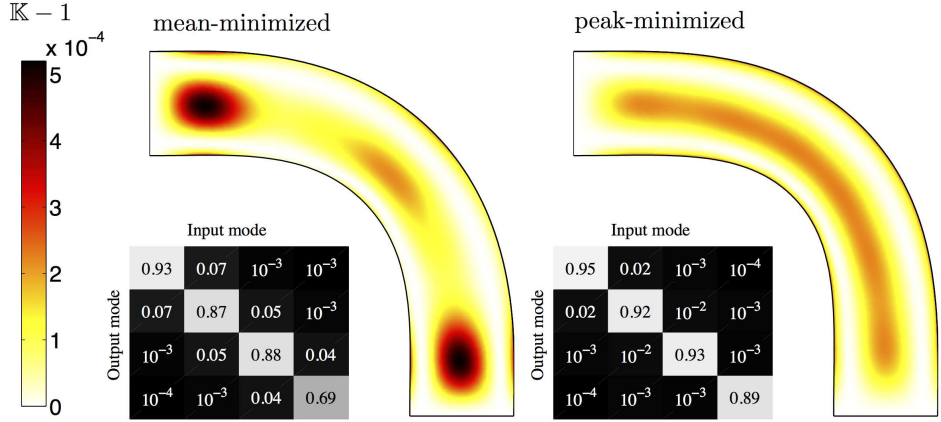}}\caption{Anisotropy profile and scattered-power matrices for optimized designs
that minimize the mean and the peak, with $R=2.5$, $N_{\ell}=3$,
and $N_{m}=6$. \label{fig:mean_max}}
\end{figure}

We found a clear difference between minimizing the peak anisotropy
versus minimizing the mean. The results of an optimization run with
$R=2.5$, $N_{\ell}=3$, and $N_{m}=6$ are shown in Fig.~\ref{fig:mean_max}.
Both structures had very low mean anisotropy~$\left\langle \mathbb{K}\right\rangle _{\mathbf{x}}-1$.
The mean-minimized structure, at $\left\langle \mathbb{K}\right\rangle -1\approx10^{-5}$,
had a slightly lower mean than the peak-minimized structure which
had $\left\langle \mathbb{K}\right\rangle -1\approx1.5\times10^{-5}$.
However, in terms of the peak anisotropy, the peak-minimized structure
is the clear winner by a factor of $2.5$, with $\max_{\mathbf{x}}\mathbb{K}-1\approx2\times10^{-4}$
as opposed to $\max_{\mathbf{x}}\mathcal{\mathbb{K}}-1\approx5\times10^{-4}$
for the mean-optimized structure. Both structures were scalarized
and tested in finite-element Maxwell simulations of the four lowest-frequency
modes of the straight waveguide. The scattered-power matrix shows
that the difference in $\max_{\mathbf{x}}\mathbb{K}$ resulted in
an order of magnitude reduction in the intermodal scattering (as shown
in the off-diagonal elements) and noticeably improved transmission,
(especially in the element $T_{44}=0.89$ for the fourth mode).

\subsection{Tradeoff between anisotropy and radius\label{sub:Tradeoff-between-anisotropy}}

\begin{figure}[t]
\centerline{\includegraphics[scale=0.15]{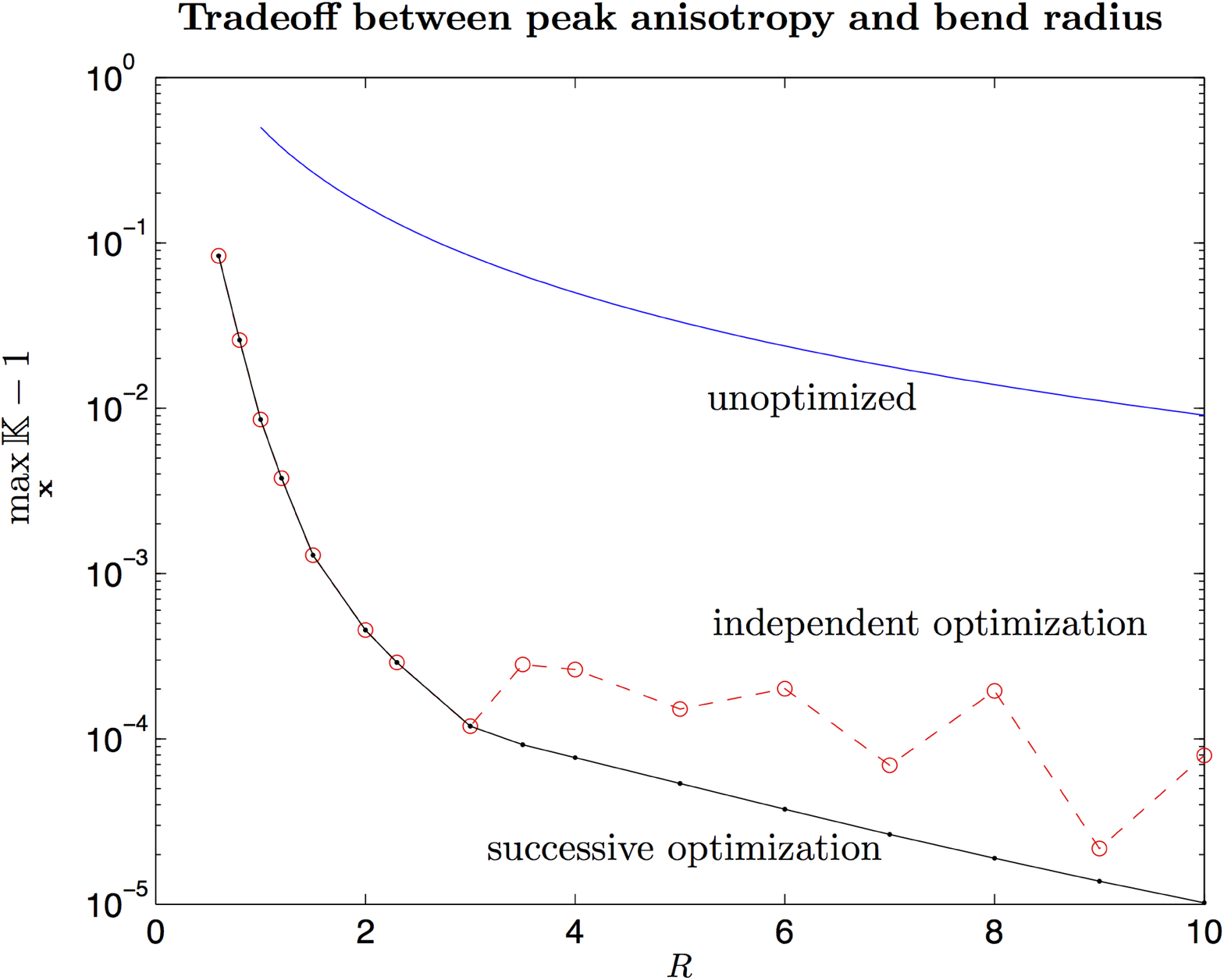}}\caption{Successive optimization with $N_{\ell}=5$, $N_{m}=8$ results in
a power law decaying tradeoff~$\max_{\mathbf{x}}\mathbb{K}-1\sim R^{-4}$
at low $R$ and an exponentially decaying tradeoff at higher $R$.
For comparison, the unoptimized anisotropy for the circular TO bend
is shown above. \label{fig:tradeoff}}
\end{figure}

In optimized structures, we found that $\max_{\mathbf{x}}\mathcal{\mathbb{K}}$
for the optimized bend, similar to the circular TO bend, decreases
monotonically with $R$ (as shown in Fig.~\ref{fig:tradeoff}). Unlike
the circular bend, however, this tradeoff seems asymptotically \emph{exponential
}rather than $\mbox{\ensuremath{\mathcal{O}}}(R^{-2})$. In particular,
there are two clearly different regimes for this tradeoff: a power
law~$\mathbb{K}-1\sim R^{-4}$ at small $R\lesssim3$ and an exponential
decay~$\mathbb{K}-1\sim\exp(-0.34R)$, at larger $R$. The second
regime was only attained after using successive optimization, because
with only one independent optimization run the algorithm tended to
get stuck in local minima. For successive optimization, the optimum
structure is used as a starting guess for the next run, and the initial
step size is set large enough so that the algorithm can reach better
local minima than the previous one.

For $R\lesssim3$, we found that there are multiple local minima and
that independent optimizations for different $R$ tend to be trapped
in suboptimal local minima, as shown by the open dots in Fig.~\ref{fig:tradeoff}.
To avoid this problem, we used a ``successive optimization'' technique
in which the optimal structure for smaller $R$ is rescaled as the
starting guess for local optima at a larger $R$, in order to stay
along the exponential-tradeoff curve. (Another possible heuristic
is ``successive refinement'' \cite{ref-robust,ref-robust_design,ref-largecircuit,ref-fast_block},
in which optima for smaller $N_{\ell,\, m}$ are used as starting
points for optimizing using larger $N_{\ell,\, m}$.)

\section{Mode squeezer\label{sec:Mode-squeezer}}

We also applied transformation inverse design to another interesting
geometry: a mode squeezer that concentrates modes and their power
in a small region in space, again with minimal intermodal scattering
(quite unlike a conventional lens, which is intrinsically angle/mode-dependent),
similar to the problem considered in \cite{ref-squeezing}
(which did not construct isotropic designs). We choose the untransformed
region to be $-1\le x\le1$ and $0\le y\le L$. The goal of this transformation~$\mathbf{x}^{\prime}(\mathbf{x})$
is to focus the beam by minimizing the \emph{mid}-\emph{beam width}
\[
W\equiv\intop_{-1}^{1}\mathrm{d}x\,\left.\sqrt{\left(\frac{\partial x^{\prime}}{\partial x}\right)^{2}+\left(\frac{\partial y^{\prime}}{\partial x}\right)^{2}}\right|_{y=\frac{L}{2}}.
\]
As in Sec.~\ref{sub:Spectral-parameterization}, the transformation
is written as a perturbation from the identity transformation (which
was used as the starting guess) and parameterized in the spectral
basis 
\begin{align*}
x^{\prime}(x,\, y) & =x+\sum_{\ell,\, m}^{N_{\ell},\, N_{m}}C_{\ell m}^{x}T_{\ell}(x)\,\sin\frac{\left(2m+1\right)\pi y}{L}\\
y^{\prime}(x,\, y) & =y+\sum_{\ell,\, m}^{N_{\ell},\, N_{m}}C_{\ell m}^{y}T_{\ell}(x)\,\sin\frac{\left(2m+1\right)\pi y}{L},
\end{align*}
The sine series automatically satisfies mirror symmetry about $y=\frac{L}{2}$
and continuity of $\mathbf{x}^{\prime}$ at the input/output interfaces~$y=0,\, L$.
However, we found that constraining the coefficients~$C^{x,\, y}$
to enforce continuity of $\mathcal{J}$ (as in Sec.~\ref{sub:Spectral-parameterization})
was not necessary (although it might give a better result) since the
optimization algorithm only squeezed the center region while leaving
the interfaces and the regions around them relatively untouched. In
this problem, we could either minimize $\mathbb{K}$ for a fixed $W$
or minimize $W$ for a fixed $\mathbb{K}$, and we happened to choose
the latter. 

\begin{figure}[H]
\centerline{\includegraphics[scale=0.15]{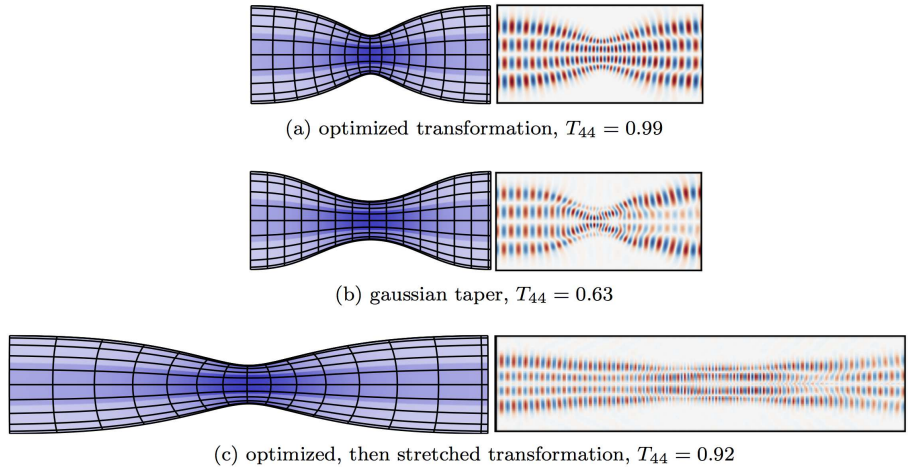}}

\caption{Optimized squeezer outperforms gaussian taper and stretched optimized
squeezers in finite element simulations. \label{fig:squeezer}}
\end{figure}

Finite-element Maxwell simulations, shown in Fig.~\ref{fig:squeezer},
demonstrate that the optimized design is greatly superior to a simple
Gaussian taper transformation designed by hand. The Gaussian transformation
was given by $x^{\prime}(\mathbf{x})=x-x\alpha\exp[-\beta(y-\frac{L}{2})^{2}]$,
where $\beta>0$ and $0<\alpha<1$. Superficially, the design seems
similar to an ``adiabatic'' taper between a wide low-index waveguide
and a narrow high-index waveguide, and it is known that any sufficiently
gradual taper of this form would have low scattering due to the adiabatic
theorem\cite{ref-sgj-coupledmode}. However, the optimized TO design
is much too short to be in this adiabatic regime. If it were in the
adiabatic regime, then taking the same design and simply stretching
the index profile to be more gradual (a taper twice as long) would
reduce the scattering, but in Fig.~\ref{fig:squeezer} we perform
precisely this experiment and find that the stretched design \emph{increases}
the scattering.

\section{Concluding remarks}

The analytical simplicity of TO design---no Maxwell equations need
be solved in order to warp light in a prescribed way---paradoxically
makes the application of computational techniques more attractive
in order to discover the best transformation by rapidly searching
a large space of possibilities. Previous work on TO design used optimization
to some extent, but overconstrained the transformation by fixing the
boundary shape while underconstraining the boundary conditions required
for reflectionless interfaces. In fact, even our present work imposes
more constraints than are strictly necessary---as long as we require
continuous $\mathcal{J}$ at the input/output interfaces, there is
no conceptual reason why those interfaces need be flat. A better bend,
for example, might be designed by constraining the location of only
two corners (to fix the bend radius) and constraining only $\mathcal{J}$
on other parts of the endfacets. However, we already achieve an exponential
tradeoff between radius and anisotropy, so we suspect that further
relaxing the constraints would only gain a small constant factor rather
than yielding an asymptotically faster tradeoff. In the case of the
mode squeezer, one could certainly achieve better results by imposing
the proper $\mathcal{J}$ constraints at the endfacets. It would also
be interesting to apply similar techniques to ground-plane cloaking.

All TO techniques suffer from some limitations that should be kept
in mind. First, TO seems poorly suited for optical devices in which
one \emph{wants} to discriminate between modes (e.g. a modal filter)
or to scatter light between modes (e.g. a mode transformer). TO is
ideal for devices in which it is desirable that all modes be transported
equally, with no scattering. Even for the latter case (such as our
multimode bend), however, TO designs almost certainly trade off computational
convenience for optimality, because they impose a stronger constraint
than is strictly required: TO is restricted to designs where the solutions
at \emph{all points} in the design are coordinate transformations
of the original system, whereas most devices are only concerned with
the solutions at the endfacets. For example, it is conceivable that
a more compact multimode bend could be designed by allowing intermodal
scattering\emph{ within} the bend as long as the modes scatter back
to their original configurations by the endfacet; the interior of
the bend might not even be a waveguide, and instead might be a resonant
cavity of some sort\cite{ref-hightrans,ref-highdensity,ref-rightangle,ref-zbend}.
However, optimizing over such structures seems to require solving
Maxwell's equations in some form at each optimization step, which
is far more computationally expensive than the TO design and, unlike
the TO design, must be repeated for different wavelengths and waveguide
designs.

\section*{Appendix}

In this Appendix, we briefly derive the fact, mentioned in Sec.~\ref{sub:Scalarization-errors-for},
that the endfacet discontinuity scales much worse with anisotropy
than the scalarization errors in the transformation interior, which
leads us to impose an explicit continuity constraint on the Jacobian
$\mathcal{J}$. In particular, we examine the Jacobian $\mathcal{J}$
for nearly isotropic transformations ($\mathbb{K}\approx1$) that
also have $\mathbf{x}^{\prime}=\mathbf{x}$ explicitly constrained
at the interfaces. (The following analysis can also be straightfowardly
extended to situations where $\mathbf{x}^{\prime}$ is a simple rotation
of $\mathbf{x}$ on the interface, or where the interface has an arbitrary
shape.) In this case, the Jacobian is
\[
\mathcal{J}=\begin{pmatrix}1 & 0\\
\delta & 1+\Delta
\end{pmatrix},
\]
where $\delta\equiv\frac{\partial x^{\prime}}{\partial y}$ and $\Delta\equiv\frac{\partial y^{\prime}}{\partial y}-1$
are small quantities ($\ll1$) if $\mathcal{J}^{T}\hspace{-2bp}\mathcal{J}$
is nearly isotropic. The anisotropy Eq.~(\ref{eq:kminusone}) is
then: 
\begin{align}
\mathbb{K}-1 & =\frac{1+\delta^{2}+\left(1+\Delta\right)^{2}}{2\left(1+\Delta\right)}-1\nonumber \\
 & =\frac{1}{2}\left(\delta^{2}+\Delta^{2}\right)+\mathcal{O}\left(\delta^{2}\Delta+\Delta^{3}\right).\label{eq:k-1_delta}
\end{align}
The determinant then satisfies
\begin{align*}
\det\mathcal{J}-1 & =\Delta\\
 & =\sqrt{2\left(\mathbb{K}-1\right)-\delta^{2}}+\mathcal{O}(\delta^{2}\Delta+\Delta^{3})\\
 & =\mathcal{O}(\sqrt{\mathbb{K}-1}).
\end{align*}
This square-root dependence is also reflected in the refractive index~$n^{\prime}=\sqrt{\varepsilon^{\prime}\mu^{\prime}}$
and leads to $\mathcal{O}(\mathbb{K}-1)$ power loss due to interface
reflections that overwhelm the $\mathcal{O}[\left(\mathbb{K}-1\right)^{2}]$
corrections to scattered power due to the scalarization of nearly
isotropic transformations (as explained in Sec.~\ref{sub:Scalarization-errors-for}).
Hence, it becomes necessary to explicitly constrain $\mathcal{J}=\mathbb{I}$
\emph{in addition }to $\mathbf{x}^{\prime}=\mathbf{x}$.

\section*{Acknowledgment}

This work was supported in part by the AFOSR MURI for Complex and
Robust On-chip Nanophotonics (Dr. Gernot Pomrenke), grant number FA9550-09-1-0704.
 
\end{document}